\documentclass[a4paper]{article}
\begin {document}
\title {\bf PROBABILITY CURRENT AND TRAJECTORY REPRESENTATION}
\author {A. Bouda \\
Laboratoire de Physique Th\'eorique, Universit\'e de B\'eja\"\i a,\\ 
Route Targa Ouazemour, 06000 B\'eja\"\i a, Alg\'erie,\\
e-mail: bouda\_a@yahoo.fr}
\date{}

\maketitle

\vskip\baselineskip
\vskip\baselineskip

\begin{abstract}
\noindent
A unified form for real and complex wave functions is proposed for 
the stationary case, and the quantum Hamilton-Jacobi equation is 
derived in the three-dimensional space. The difficulties which appear in 
Bohm's theory like the vanishing value of the conjugate momentum in 
the real wave function case are surmounted. In one dimension, a new 
form of the general solution of the quantum Hamilton-Jacobi equation
leading straightforwardly to the general form of the Schr\"odinger 
wave function is proposed. For unbound states, it is shown that 
the invariance of the reduced action under a dilatation plus a rotation 
of the wave function in the complex space implies that microstates do 
not appear. For bound states, it is shown that some freedom subsists 
and gives rise to the manifestation of microstates not detected by 
the Schr\"odinger wave function.
\end{abstract}

\vskip\baselineskip


\vskip0.5\baselineskip
\vskip0.5\baselineskip

\noindent
{\bf 1\ \ INTRODUCTION }

\vskip0.5\baselineskip

The debate opened by Einstein and Bohr about the interpretation of
quantum mechanics is far from being closed. Among all attempts to
obtain a deterministic theory, the approach proposed by Bohm \cite{Bohm}
is one of the most interesting. The starting point is the Schr\"odinger
equation
\begin {equation}
 - {\hbar^2 \over {2m}  } \Delta\psi + V\psi = i\hbar  
 {\partial\psi \over \partial t  }\ , 
\end {equation} 
which describes the evolution of the wave function of a
non-relativistic spinless particle of mass $m$ in a potential
$V$. Bohm writes the wave function in the form
\begin {equation}
\psi(x,y,z,t) = A(x,y,z,t)\  
\exp\left({i\over\hbar }S(x,y,z,t)\right)\ ,
\end {equation}
where $A(x,y,z,t)$ and $S(x,y,z,t)$ are real functions. By
substituting (2) in (1), it follows
\begin {equation}
{1\over 2m }(\vec{\nabla} S)^2- {\hbar^2\over 2m }{\Delta A\over A }+V=
-{\partial S \over \partial t  }\ ,
\end {equation}
\begin {equation}
 \vec{\nabla  }\cdot \left(A^2 {\vec{\nabla  } S\over m} \right) = 
-{\partial {A^2} \over \partial t  } \ . 
\end {equation}
The term proportional to $\hbar^2 $ in Eq. (3)
\begin {equation}
 V_B \equiv -{\hbar^2\over {2m} }{\Delta A\over A } 
\end {equation}
is called the Bohm quantum potential. At first sight, setting
$\hbar= 0$ in Eq. (3), making $V_B$ vanish, gives the classical 
Hamilton-Jacobi equation which describes the motion of the particle. 
$S$ is then identified as the reduced action and $V_B$ is interpreted 
as describing the quantum effects. However, the classical limit 
$\hbar\to 0$ is not trivial. In fact, Floyd \cite{F1} showed 
that, in general, when taking the limit $\hbar\to 0$, a residual 
indeterminacy subsists.  

Relation (4) represents the conservation equation of the probability
current. Indeed, if one substitutes (2) in the expression of the current
\begin {equation}
\vec{\jmath}={\hbar\over 2mi }(\psi^*\vec{\nabla}{\psi}-
\psi\vec{\nabla}{\psi^*})\ , 
\end {equation}
one finds
\begin {equation}
\vec{\jmath}=A^2{\vec{\nabla} S\over m}\ .  
\end {equation}
This expression is a product of the probability density
$\vert\psi\vert^2=A^2$ by $ \vec{v} \equiv \vec{\nabla} S/ m $ 
which was recognized by Bohm \cite{Bohm} and de Broglie \cite{LdB}
in his pilot wave theory as the velocity of the particle. In the
stationary case, where
\begin {equation}
 S(x,y,z,t)= S_0(x,y,z)-Et\ , 
\end {equation}
\begin {equation}
{\partial A \over \partial t  }(x,y,z,t) =0\ ,   
\end {equation}
and the constant $E$ representing the energy of the particle, Floyd
\cite{F2} showed that in one-dimensional space, the velocity was
not given by $\;m^{-1} \partial S_0/\partial x \;$ as defined by Bohm 
\cite{Bohm} and de Broglie \cite{LdB}, but by the expression
\begin {equation}
{dx\over dt}={{\partial S_0 / \partial x }\over {m(1-
\partial V_B /\partial E}) }\ .  
\end {equation}

The reduced action $S_0$ as defined by (2) and (8) cannot be used
to define correctly the conjugate momentum as $\vec{\nabla} S_0$. To 
see this, consider the case in which the wave function is real as it is
for the ground state of hydrogeno\"\i d atoms or the one-dimensional
harmonic oscillator. In this case, using (2), $S_0$ is constant
and then the conjugate momentum has a vanishing value. Obviously, this
conclusion is not satisfactory.

In the one-dimensional case, Floyd \cite{F2,F3} surmounted this difficulty by 
using a trigonometric representation. He showed then that trajectory 
representation described microstates \cite{F2,F4,F5} not detected by the 
Schr\"odinger wave function. In a recent paper \cite{F6}, the same 
author synthesized the basic ideas of trajectory representation 
and contrasted them to those of the Copenhagen school and those of 
Bohmian mechanics.

Recently, by assuming that all quantum systems can be connected by a
coordinate transformation, Faraggi and Matone \cite{FM1,FM2} have 
derived in one dimension the quantum Hamilton-Jacobi equation, which
in turn, leads to the Schr\"odinger equation. In other words, from an
equivalence principle, they deduced quantum mechanics. In particular,
without using the usual axiomatic interpretation of the wave 
function, they showed that the tunnel effect and the energy 
quantization are consequences of the equivalence principle 
\cite{FM2,FM3}. These authors, together with Bertoldi \cite{BFM}, 
deduced the higher dimensions version of the quantum Hamilton-Jacobi 
equation. They showed in one dimension that there are microstates. 
They deduced also in higher dimensions the existence of a hidden 
antisymmetric two-tensor field which can play an important role in the 
search for the quantum origin of fundamental interactions. This 
connection between the equivalence principle and quantum mechanics 
gives some hope for a unified description of fundamental interactions. 
Furthermore, Matone \cite{Ma} has suggested that gravity interaction 
can be derived from the quantum potential.

In this paper, we reproduce in three dimensions some results found in 
Ref. \cite{BFM} without appealing to differential geometry. We also give
some consequences of the approach presented here about microstates.
In section 2, a relationship between the wave function and the reduced
action is constructed. The probability current is used in 
section 3 to establish the quantum Hamilton-Jacobi equation for the 
complex wave functions and in section 4 for the real wave functions. 
In section 5, a new form of the general solution of the quantum 
Hamilton-Jacobi equation in one dimension is proposed. It is shown 
that there is no trace of microstates for unbound states. However, 
some freedom subsists for bound states giving rise to the 
manifestation of microstates. In three dimensions, the hidden 
antisymmetric two-tensor field is presented. Section 6 is devoted 
to conclusions.

\vskip0.5\baselineskip
\vskip0.5\baselineskip

\noindent
{\bf 2\ \ THE FORM OF THE WAVE FUNCTION}

\vskip0.5\baselineskip

Let us begin by the following remark. If one sets
\begin {equation}
\psi(x,y,z,t) = A(x,y,z,t)\; \exp\left(-{i\over\hbar }S(x,y,z,t)\right)
\end {equation}
and substitutes this expression in the Schr\"odinger equation,
one gets
\begin {equation}
{1\over 2m }(\vec{\nabla} S)^2- {\hbar^2\over 2m }{\Delta A\over A }+V=
{\partial S \over \partial t  }\ , 
\end {equation}
\begin {equation}
 \vec{\nabla  }\cdot \left(A^2 {\vec{\nabla  } S\over m} \right) = 
{\partial A^2 \over \partial t  }\ .
\end {equation}
In a non-stationary case $(\partial A/ \partial t \not= 0)$,
by comparing Eqs. (12) and (13) with (3) and (4), it is easy to deduce 
that expressions (2) and (11) cannot be simultaneously
solution of the Schr\"odinger equation. In the stationary case, the
situation is different. In fact, if one replaces successively the two
expressions
\begin {equation}
\psi_1 = \exp\left(-{i\over\hbar }Et\right)A\; 
\exp\left({i\over\hbar }S_0\right)\ ,  
\end {equation}
\begin {equation}
\psi_2 = \exp\left(-{i\over\hbar }Et\right) A\;
\exp\left(-{i\over\hbar }S_0\right)  
\end {equation}
in the Schr\"odinger equation, one gets the same equations
\begin {equation}
{1\over 2m }(\vec{\nabla} S_0)^2- {\hbar^2\over 2m }{\Delta A\over A }+V=
E \ , 
\end {equation}
\begin {equation}
 \vec{\nabla  }\cdot (A^2 \vec{\nabla  } S_0) = 0 \ . 
\end {equation}
This means that if $\psi_1$ (respectively $\psi_2$) is solution of the
Schr\"odinger equation, $\psi_2$ (respectively $\psi_1$) is also
solution. It follows that for the physical wave function in the 
stationary case we can choose the form
\begin {equation}
\psi(x,y,z,t) = \exp\left(-{i\over\hbar }Et\right)\; \phi(x,y,z) \ ,
\end {equation}
in which $\phi(x,y,z)$ is the linear combination
\begin {equation}
 \phi(x,y,z)= A(x,y,z)\left[\alpha\ \exp\left({i\over\hbar }S_0(x,y,z)\right)
+\beta\ \exp\left(-{i\over\hbar }S_0(x,y,z)\right)\right] \ ,  
\end {equation}
$\alpha $ and $\beta$ being complex constants. Note that this form (19)
of the wave function has been established with a different method in 
one dimension \cite{FM1,FM2} and in higher dimensions \cite{BFM}.

Now, if one replaces in the Schr\"odinger equation $\psi$ by the 
expressions given in (18) and (19), one finds
\begin{eqnarray}
\left[{1\over 2m }(\vec{\nabla} S_0)^2 
- {\hbar^2\over 2m }{\Delta A\over A }
+ V - E\right]
\left[\alpha\ \exp\left({i\over\hbar }S_0\right)
+\beta\ \exp\left(-{i\over\hbar }S_0\right)\right]- && 
\nonumber\\
 {i\hbar\over 2mA^2 }\vec{\nabla  }\cdot (A^2 \vec{\nabla  } S_0)
\left[\alpha\ \exp\left({i\over\hbar }S_0\right)
-\beta\ \exp\left(-{i\over\hbar }S_0\right)\right]= 0 \ .&&
\end{eqnarray}
Before analyzing the content of this equation, let us calculate the
probability current. If one replaces $(18)$ in $(6)$, one gets
\begin {equation}
\vec{\jmath}=({\vert{\alpha}\vert}^2-{\vert{\beta}\vert}^2)A^{2}
{\vec{\nabla} S_0\over m} \ .
\end {equation}
This form of the current will play a crucial role in the approach
which is developed here.

\vskip0.5\baselineskip
\vskip0.5\baselineskip

\noindent
{\bf 3\ \ THE COMPLEX WAVE FUNCTION}

\vskip0.5\baselineskip

In what follows, one should understand by real wave function, any
function which can be written as a product of a constant, which could
be complex, with a real function.

In order to show that the wave function (19) cannot be real when
$\vert\alpha\vert \not=\vert\beta\vert$, let us set
\begin {equation}
\alpha =\vert\alpha\vert \ \exp (ia) \ ,\ \ \ \ \ \ \ \ \ \ 
\beta =\vert\beta\vert\ \exp (ib) \ ,
\end {equation}
with $a$ and $b$ real constants. Expression (19) can then be 
written in the form
\begin {eqnarray}
\phi = A\ \exp\left(i{{a+b}\over 2 }\right)\
\left[ (\vert\alpha\vert + \vert\beta\vert)\
\cos\left( {S_0\over\hbar } + {{a-b}\over 2} \right)\right.\ +\hskip15mm&& 
\nonumber\\
 \left. i(\vert\alpha\vert - \vert\beta\vert)\
\sin\left( {S_0\over\hbar } + {{a-b}\over 2} \right) \right] \ .  && 
\end {eqnarray}

\noindent 
Knowing that $S_0$ is a function of $(x,y,z)$, this last
expression shows clearly that when $\vert\alpha\vert \not=
\vert\beta\vert$, the wave function cannot be brought back to a
product of a constant by a real function.

Now, to derive the quantum Hamilton-Jacobi equation, let us use
expression (21) for the probability current. The conservation
equation, which is a consequence of the Schr\"odinger equation, can be
written as
\begin {equation}
\vec{\nabla}\cdot \left[(\vert{\alpha}\vert^2-\vert{\beta}\vert^2)A^2
{\vec{\nabla} S_0\over m}\right]=0 \ . 
\end {equation}
Therefore, for the complex wave functions $(\vert\alpha\vert \not=
\vert\beta\vert)$, Eq. (24) turns out to be
\begin {equation}
\vec{\nabla}\cdot (A^2\vec{\nabla} S_0)=0 \ .  
\end {equation}
Eq. (20) reduces then to
\begin {equation}
{1\over 2m }(\vec{\nabla} S_0)^2- {\hbar^2\over 2m }{\Delta A\over A }
+V-E =0 \ .
\end {equation}
Although the last two equations have the same form as those
established by Bohm \cite{Bohm}, they are fundamentally different. 
In fact, in Bohm's theory, the physical solution of the Schr\"odinger
equation is written in the form (14). In this case, Eqs. (16) and 
(17) represent the Bohm's equations and then the functions $A$ and
$S_0$ appearing in (16) and (17) are not the same as those of Eqs. 
(25) and (26). The reason is that in Bohm's theory, Eqs. (16) and 
(17) are derived by writing the physical solution in the form 
(14) which does not allow to define correctly the conjugate momentum, 
while in the present approach Eqs. (25) and (26) are obtained by 
writing the same physical solution in the form (19).

\vskip0.5\baselineskip
\vskip0.5\baselineskip

\noindent
{\bf 4\ \ THE REAL WAVE FUNCTION}

\vskip0.5\baselineskip

In the case where $\vert\alpha\vert =\vert\beta\vert$, and using
Eq. (23), the physical wave function defined by (19) becomes
\begin {equation}
\phi = 2\vert\alpha\vert A\  \exp\left( i{a+b\over 2 } \right)
\cos\left( {S_0\over\hbar } + {a-b\over 2} \right)\ .
\end {equation}
It is clear that the wave function is real up to a constant phase
factor.

Here the vanishing of the probability current is expressed by the fact
that $\vert\alpha\vert =\vert\beta\vert$, and not by 
$\vec{\nabla} S_0=\vec{0}$ as in the case of Bohm's approach.

Using (22) with $\vert\alpha\vert =\vert\beta\vert$, Eq. (20) 
turns out to be
\begin {equation}
{1\over 2m }(\vec{\nabla} S_0)^2- {\hbar^2\over 2m }{\Delta A\over A }
+V + {\hbar\over 2mA^2 }\left[\vec{\nabla  }\cdot (A^2 \vec{\nabla  } S_0)
\right]\ \tan \left({S_0\over \hbar}+{a-b\over 2} \right) = 
E\ . 
\end {equation}
Comparing with the usual quantum Hamilton-Jacobi equation, (28)
contains an additional term proportional to $\hbar$.

At first glance, one may think that for any function $\phi(x,y,z)$
describing a physical state, there is an infinite number of ways to
choose the couple $(A,S_0)$ in such a way as to satisfy relation
(27). For example, if one chooses $S_0$ to be constant, Eq. (28)
becomes
\begin {equation}
-{\hbar^2\over 2m }{\Delta A\over A }+V=E 
\end {equation}
which is exactly the Schr\"odinger equation. Another possible choice
is to take $A=cst.$ and deduce the equation
\begin {equation}
{1\over 2m }(\vec{\nabla} S_0)^2 + V + {\hbar\over 2m }\Delta S_0\ 
\tan \left({S_0\over \hbar}+{a-b\over 2} \right) = E  
\end {equation}
from which one can reproduce the Schr\"odinger equation.

Among all these choices, is there any couple $(A,S_0)$ in which $S_0$
is the good function defining correctly the conjugate momentum by
$\vec{\nabla} S_0$ ? In other words, is there any particular relation
between $A$ and $S_0$ ?

To answer this crucial question, let us analyze the physics content 
of expression (21) for the probability current. This expression
suggests that $\;\vec{\jmath}\;$ is a sum of two currents
\begin {equation}
\vec{\jmath}= \vec{\jmath}_+ + \vec{\jmath}_- \ ,  
\end {equation}
where 
$\;\vec{\jmath}_+ = \vert{\alpha}\vert^{2}A^{2} \vec{\nabla} S_0 / m\;$ 
and 
$\;\vec{\jmath}_- = -\vert{\beta}\vert^{2}A^{2}\vec{\nabla} S_0 / m\;$
correspond to the two opposite directions of motion of the particle
along the trajectory.  The fact that the current has a vanishing 
value in the case of a real wave function
$(\vert\alpha\vert =\vert\beta\vert)$ 
means that there is an equal
probability to have the particle move in one direction or in the
other.

Thus, to each direction of motion along the trajectory, it is natural
to associate one of the wave functions
\begin {equation}
\phi_1 = A\ \exp\left({i\over\hbar }S_0\right)\ ,    
\end {equation}
\begin {equation}
\phi_2 =  A\ \exp\left(-{i\over\hbar }S_0\right)\ ,  
\end {equation}
which were combined in Eq. (19) to obtain expression (21)
for the current. This means that $\phi_1$ and $\phi_2$ must be
simultaneously solution of the Schr\"odinger equation. Thus, there is
no reason why this should not happen in the particular case
$\vert\alpha\vert = \vert\beta\vert$. Consequently, the couple
$(A,S_0)$ must be chosen in such a way as to impose to $\phi_1$ and
$\phi_2$ to be solutions of Schr\"odinger's equation knowing that
expression (27) is also solution. To satisfy this condition, it
is sufficient to require that the function
\begin {equation}
 \theta(x,y,z)=  A\ 
\sin\left({S_0\over\hbar }+{a-b\over 2} \right)  
\end {equation}
be a solution of Schr\"odinger's equation. In fact, if $\phi$
and $\theta$ are solutions, then $\phi_1$ and $\phi_2$ are also
solutions since they are linear combinations of $\phi$ and $\theta$.

Of course, if one substitutes (27) in the Schr\"odinger equation,
one gets (28). On the other hand, substituting $\theta$ by its 
expression (34) in the Schr\"odinger equation, one gets
\begin {equation}
{1\over 2m }(\vec{\nabla} S_0)^2- {\hbar^2\over 2m }{\Delta A\over A }
+V - {\hbar\over 2mA^2 }\left[\vec{\nabla  }\cdot (A^2 \vec{\nabla  } S_0)
\right]\ \cot \left({S_0\over \hbar}+{a-b\over 2 }\right) = E\ . 
\end {equation}
It is clear that Eqs. (28) and (35) cannot be simultaneously
satisfied unless one has
\begin {equation}
 \left[\vec{\nabla  }\cdot (A^2 \vec{\nabla  } S_0)
\right]\ \tan \left({S_0\over \hbar}+{a-b\over 2 }\right) =
- \left[\vec{\nabla  }\cdot (A^2 \vec{\nabla  } S_0)
\right]\ \cot \left({S_0\over \hbar}+{a-b\over 2 }\right)\ .
\end {equation}
This implies that either
	$\;\tan^{2} \left({S_0\over \hbar}+{a-b\over2} \right) =-1\;$ 
which is not possible, or
\begin {equation}
\vec{\nabla  }\cdot (A^2 \vec{\nabla  } S_0) =0\ .  
\end {equation}
In conclusion, the couple $(A,S_0)$ must be chosen in such a way as to
satisfy Eq. (37). However, as we will see in the next section, for any 
physical state $\phi$, relations (27) and (37) do not always fix 
univocally $A$ and $S_0$. In other words, there are some cases where  
some freedom subsists in the choice of $A$ and $S_0$. Eq. (37), 
imposed by physical considerations, implies that (28) reduces to
\begin {equation}
{1\over 2m }(\vec{\nabla} S_0)^2- {\hbar^2\over 2m }{\Delta A\over A }
+V = E\ . 
\end {equation}
Eqs. (37) and (38) are exactly the same as those obtained for
the complex wave functions in the last section.

Thus, for both real and complex wave functions, we obtain the same
quantum Hamilton-Jacobi equation (26) or (38), and the functions
$A$ and $S_0$ are related by the same equation (25) or (37).

As shown by Bertoldi-Faraggi-Matone \cite{BFM}, these results can
be also obtained within the framework of differential geometry 
in the following way. From the equivalence principle, which
stipulates that physical states are equivalent under coordinate 
transformations, one can derive the cocycle condition in three 
dimensions. Then, without appealing to Schr\"odinger's equation, 
this condition leads to the quantum Hamilton-Jacobi equation.

\vskip0.5\baselineskip
\vskip0.5\baselineskip

\noindent
{\bf 5\ \ MICROSTATES, HIDDEN ANTISYMMETRIC TENSOR \\ 
AND VELOCITY}

\vskip0.5\baselineskip

\vskip0.5\baselineskip
\noindent
{\bf 5.1\ \ The quantum Hamilton-Jacobi equation in one dimension}
\vskip0.5\baselineskip

For both real and complex wave functions, one can integrate (25) or 
(37) to obtain
\begin {equation}
A=k\left({\partial S_0 \over \partial x  }\right)^{-1/2  }\ ,
\end {equation}
where $k$ is a constant of integration. Then, by substituting this
expression in (26) or (38), we obtain the well-known equation
\cite{Mess}
\begin {equation}
{1\over 2m} \left({\partial S_0 \over \partial x}\right)^2- 
{\hbar^2\over 4m}  \left[{3\over 2}\left( 
{\partial S_0 \over\partial x}\right)
^{- 2 }\left({\partial^2 S_0 \over \partial x^2}\right)^2-
\left( {\partial S_0 \over \partial x  }\right)^{- 1 }
\left({\partial^3 S_0 \over \partial x^3  }\right) \right]+V=E\ .
\end {equation}
Of course, this equation is different from the usual one because the
function $S_0$ which appears here is related to the Schr\"odinger 
wave function by (19). Note that it is not possible to obtain such 
an equation for the real wave functions in Bohm's theory.

Before going further, it is interesting to see what the difference 
is between classical mechanics and quantum mechanics. Setting 
$\hbar= 0$ in (40), we obtain the classical Hamilton-Jacobi equation. 
It is a differential equation of first-order and its solution
$S_0^{cl}=S_0^{cl}(x,E)$
depends on the only one non-additive integration constant $E$ which 
can be determined by the usual initial conditions. When we take  
into account the quantum effects $(\hbar\not= 0)$, (40) is a 
differential equation of third-order and its solution 
$S_0^{Q}=S_0^{Q}(x,E,l_1,l_2)$
contains two another non-additive integration constants $l_1$ and $l_2$
which can be determined by further initial conditions 
\cite{F2,F7}. Thus, $l_1$ and $l_2$ can be considered as
hidden variables.

\vskip0.5\baselineskip
\noindent
{\bf 5.2\ \ The solution}
\vskip0.5\baselineskip

In this subsection, we propose a new form of the general solution of the 
quantum Hamilton-Jacobi equation and show that this solution
leads straightforwardly to the form (19) of the wave function. In fact,
one can check that the expression
\begin {equation}
S_0=\hbar \ \arctan{\left({\sigma \theta_1 + \nu \theta_2} \over 
{\mu \theta_1 + \gamma \theta_2}\right)} + \hbar \lambda \ ,
\end {equation}
is solution of Eq. (40). The set $(\theta_1,\theta_2)$ represents 
two real independent solutions of the one dimensional stationary 
Schr\"odinger equation 
$-\hbar^2 \phi^{''} /2m + V\phi=E\phi\, $ 
and $(\mu,\nu,\sigma,\gamma,\lambda)$ are arbitrary real constants 
satisfying the condition $\mu\nu \not= \sigma\gamma $. Note that one 
of the two couples $(\mu,\nu)$ or $(\sigma,\gamma)$ can be absorbed 
by rescaling the solutions $\theta_1$ and $\theta_2$ since the 
Schr\"odinger equation is linear. Therefore, in what follows, 
we set $\sigma = \gamma =1$ and interpret $\mu$ and $\nu$ 
$(\mu\nu \not= 1)$ as integration constants of the second-order 
differential equation (40) with respect to $\partial S_0/\partial x$ 
since this derivative depends only on $\mu$ and $\nu$ and not on 
$\lambda$ 
\begin {equation}
{\partial S_0 \over \partial x }= {\hbar W(\mu\nu - 1)  \over A^2 } 
\ . 
\end {equation}
Here, 
$W = \theta_1\theta_2^{'} - \theta_2\theta_1^{'}$ 
is the Wronskian and the function $A$ is given by
\begin {equation}
A=\sqrt{(\mu^{2}+1) \theta_1^{2}+(\nu^{2}+1)\theta_2^{2} +  
2(\mu + \nu)\theta_1\theta_2} \ .
\end {equation}
The parameter $\lambda$ is an additive integration constant arising 
by integrating $\partial S_0 / \partial x$. With these three 
integration constants $(\mu, \nu, \lambda)$, expression (41)  
with $\sigma = \gamma = 1$, represents the general solution of 
(40). Note that Eq. (40) and its solution were also investigated by 
Floyd \cite{F4,F5} and by Faraggi-Matone \cite{FM1,FM2}.

Let us show now that the solution (41) leads to the general form
(19) of the wave function. From Eq. (41), we can deduce  
\begin {equation}
\theta_1 + \nu \theta_2 = A\ \sin\left({S_0\over \hbar }-\lambda\right) \ ,
\end {equation}
\begin {equation}
\mu \theta_1 + \theta_2 = A\ \cos\left({S_0\over \hbar }-\lambda\right) \ .
\end {equation}
If we solve this system for $\theta_1$ and $\theta_2$, we easily obtain
\begin {equation}
\theta_1 = A\ {\nu\ \cos\left({S_0\over \hbar }-\lambda\right)
-\sin \left({S_0\over \hbar }-\lambda\right) \over \mu \nu -1 }
\end {equation}
\begin {equation}
\theta_2 = A\ {\mu\ \sin\ \left({S_0\over \hbar }-\lambda\right) -
\cos\left({S_0\over \hbar }-\lambda\right) \over \mu \nu -1 }
\end {equation}
The general solution of the Schr\"odinger equation is given by
\begin {equation}
\phi=C_1 \theta_1+C_2\theta_2 \ .
\end {equation}
The constants $C_1$ and $C_2$ are generally complex and determined by
the initial conditions \cite{F5}
\begin {equation}
\phi(x_0)=C_1 \theta_1 (x_0)+C_2\theta_2 (x_0) \ ,
\end {equation}
\begin {equation}
\phi^{'} (x_0)=C_1 \theta_{1}^{'}(x_0)+C_2\theta_{2}^{'} (x_0) \ .
\end {equation}
If we substitute expressions (46) and (47) in the physical solution (48) 
of the Schr\"odinger equation, we obtain
\begin {equation}
 \phi= A\left[\alpha\ \exp\left({i\over\hbar }S_0\right)
+\beta\ \exp\left(-{i\over\hbar }S_0\right)\right] 
\end {equation}
in which
\begin {equation}
 \alpha =  {(\nu C_1 -C_2) + i(C_1 -\mu C_2) \over 2(\mu\nu-1) }
\exp{(-i\lambda)} \ ,
\end {equation}
\begin {equation}
 \beta =   {(\nu C_1 -C_2) - i(C_1 -\mu C_2) \over 2(\mu\nu-1) }
  \exp{(i\lambda)} \ .
\end {equation}
Expression (51) is exactly the same as the one given by Eq. (19)
and which we have used to derive the quantum Hamilton-Jacobi equation.

According to (52) and (53), it is clear that for any fixed set $(C_1,C_2)$, 
one can change arbitrarily the integration constants $\mu$, $\nu$ 
and $\lambda$ since $\alpha$ and $\beta$ are also arbitrary. 
This could then be interpreted as a symmetry of the wave function 
because $S_0$, $A$, $\alpha$ and $\beta$ defined respectively by 
(41), (43), (52) and (53) vary with $\mu$, $\nu$ and 
$\lambda$, while the wave function (51) remains invariant 
since it has been constructed from (48). At first sight, this 
symmetry would give rise to microstates.  In the next subsection, 
we will however show that this conclusion is not correct because 
in this construction we have two superfluous degrees of freedom.

\vskip0.5\baselineskip
\noindent
{\bf 5.3\ \ Microstates and unbound states}
\vskip0.5\baselineskip

In section 3, we showed that (51) could be written in the form 
(23). We can therefore write
\begin {equation}
{\rm Re}\;\left[\exp{\left(-i{{a+b} \over 2}\right)}\phi\right] = 
(\vert\alpha\vert + \vert\beta\vert)\ A\ \cos\left( {S_0\over\hbar } 
+ {{a-b}\over 2} \right) \ ,
\end {equation}
\begin {equation}
{\rm Im}\;\left[\exp{\left(-i{{a+b} \over 2}\right)}\phi \right] = 
(\vert\alpha\vert - \vert\beta\vert)\ A\ \sin\left( {S_0\over\hbar } 
+ {{a-b}\over 2} \right) \ .
\end {equation}
In the unbound states case $(\vert\alpha\vert \not= \vert\beta\vert)$,
it follows
\begin {equation}
S_0 = \hbar \ \arctan {\left({\vert\alpha\vert + \vert\beta\vert
 \over \vert\alpha\vert - \vert\beta\vert}\ 
{ {\rm \ Im}\;\left[\exp(-i(a+b)/ 2)\, \phi \right]  
\over {\rm \ Re}\ \left[\exp(-i(a+b)/ 2)\, \phi \right] }  \right)} 
+\hbar {{b-a}\over 2} \ .
\end {equation}
Before analyzing the content of this equation, let us perform the 
transformation
\begin {equation}
\alpha \to \hat{\alpha}= \Omega \alpha, \ \ \ \ \ \  \ \ \ 
     \beta \to \hat{\beta}= \Omega \beta 
\end {equation}
implying a dilatation followed by a rotation of the wave function in 
the complex space
\begin {equation}
\phi \to \hat{\phi}= \Omega \phi \ . 
\end {equation}
The parameter $\Omega=|\Omega|\ \exp{(i\omega)}$ is an arbitrary complex 
number and $\omega$ a real constant. Setting
\begin {equation}
\hat{\alpha} =\vert\hat{\alpha}\vert \ \exp (i\hat{a})\ , \ \ \ \ \ \ \ \ \ \
\hat{\beta} =\vert\hat{\beta}\vert\ \exp (i\hat{b}) \ ,
\end {equation}
we have
\begin {equation}
|\hat{\alpha}| =|\Omega||\alpha|\ , \ \ \ \ \ \ \ 
|\hat{\beta}| = |\Omega||\beta| \ \ \ \ \ \ \ \ \hat{a} = a+ \omega \ ,
\ \ \ \ \ \ \ \ \hat{b} = b+ \omega \ .  
\end {equation}
The new wave function must have the form
\begin {equation}
 \hat{\phi}= \hat{A}\left[\hat{\alpha}\ \exp\left({i\over\hbar }
\hat{S}_0\right)+\hat{\beta}\ \exp\left(-{i\over\hbar }
\hat{S}_0\right)\right]   
\end {equation}
and therefore, as in Eq. (56), the new reduced action is given by
\begin {equation}
\hat{S}_0 = \hbar \ \arctan {\left({\vert\hat{\alpha}\vert + 
\vert\hat{\beta}\vert
 \over \vert\hat{\alpha}\vert - \vert\hat{\beta}\vert}\ 
{ {\rm \ Im}\;\left[\exp(-i(\hat{a}+\hat{b})/ 2)\, \hat{\phi} \right]  
\over {\rm \ Re}\ \left[\exp(-i(\hat{a}+\hat{b})/ 2)\, \hat{\phi} \right] }  
\right)} 
+\hbar {{\hat{b}-\hat{a}}\over 2} \ .
\end {equation}
Substituting Eqs. (58) and (60) in (62), we obtain
\begin {equation}
\hat{S_0}=S_0 \ . 
\end {equation}
Now, if we substitute (57), (58) and (63) in (61) and then 
use (51), we deduce 
that
\begin {equation}
\hat{A}=A \ .
\end {equation}
In conclusion, the functions $A$ and $S_0$ are invariant under a 
dilatation and a rotation of the wave function in the complex space. 
This invariance of the reduced action allows us to perform a 
particular transformation for which $\Omega = \alpha^{-1}$ and 
identify the new wave function to the physical one given by (48)
\begin {equation}
C_1 \theta_1+C_2\theta_2 = A\left[ \exp\left({i\over\hbar}S_0\right)
+\eta\ \exp\left(-{i\over\hbar }S_0\right)\right] \ ,
\end {equation}
where $\eta = \beta /\alpha$. Since the additive integration constant 
$\lambda$ has no dynamical effects, we set $\lambda =0$ from now 
on. Therefore, substituting expressions (41) and (43) for 
$S_0$ and $A$ in the right hand side of (65), we obtain
\begin {equation}
C_1 \theta_1+C_2\theta_2 =\big[\mu+i+\eta(\mu-i)\big] \ \theta_1
 + \big[1+i\nu+\eta(1-i\nu)\big]\ \theta_2 \ .
\end {equation}
The identification of the coefficients of $\theta_1$ and $\theta_2$  
leads to
\begin {equation}
C_1 =\mu+i+\eta(\mu-i) \ , 
\end {equation}
\begin {equation}
C_2 =1+i\nu+\eta(1-i\nu) \ , 
\end {equation}
which are equivalent to Eqs. (52) and (53) if we set $\lambda=0$, 
$\alpha =1$ and $\beta = \eta$. Separating in (67) and (68) the real 
part from the imaginary part, we obtain four equations which can be 
solved with respect to 
$\,{\rm \ Re}(\eta)\,$, $\,{\rm \ Im}(\eta)\,$, $\mu$ and $\nu$. 
It follows that in this particular choice of $\Omega$, the reduced 
action $S_0$ defined by (41) with $\sigma = \gamma = 1$ is entirely 
determined. Furthermore, since $S_0$ is invariant, it keeps its value 
for any choice of $\Omega$. In conclusion, for a given physical state, 
the initial conditions (49) and (50) fix univocally the reduced action 
and therefore there is no trace of microstates for unbound states.

To clarify our point of view, we would like to make the following remark. 
Eqs. (52) and (53) can be written in the form 
\begin {equation}
C_1 =\alpha \ \Big[(\mu+i)\ \exp{(i\lambda)}\Big]+
\beta \ \Big[(\mu-i) \ \exp{(-i\lambda)} \Big]\ , 
\end {equation}
\begin {equation}
C_2 =\alpha \ \Big[(1+i\nu)\ \exp{(i\lambda)}\Big]+
\beta \ \Big[(1-i\nu)\ \exp{(-i\lambda)} \Big]\ . 
\end {equation}
These last equations show clearly that we can rescale $C_1$ and 
$C_2$ with any complex number by multiplying $\alpha$ and $\beta$ 
by the same number without any effect on $\mu$, $\nu$ and 
$\lambda$. This rescaling does not affect the wave function up to 
a multiplicative constant factor. This means that we have two 
more degrees of freedom. In others words, the fact that $\Omega$ 
can be arbitrarily chosen implies that, among the four real 
parameters defining the complex numbers $\alpha$ and $\beta$ in 
the form (51) of the wave function, we can keep only two 
parameters free. In this way, for any set $(C_1,C_2)$ of initial 
conditions, these two free parameters will be fixed for unbound 
states.  

Note that we can also arrive to the same conclusion using another 
method. In fact, substituting expression (41) for $S_0$ with  
$\sigma = \gamma =1$ in (23), we obtain
\begin {eqnarray}
\phi = A\,(\vert\alpha\vert + \vert\beta\vert)\,
\cos{\left[\arctan{\left({\theta_1 + \nu \theta_2} \over 
{\mu \theta_1 + \theta_2}\right)}  \right]}+\hskip25mm&& \nonumber\\
iA\,(\vert\alpha\vert - \vert\beta\vert)\
\sin{\left[\arctan{\left({\theta_1 + \nu \theta_2} \over 
{\mu \theta_1 + \theta_2}\right)} \right]} \ ,
\end {eqnarray}
where we have discarded the unimportant phase factor 
$\exp\left(i{{a+b}\over 2 }\right)$ and chosen the integration
constant $\lambda$ equal to $(b-a)/2$. Using expression (43) for 
$A$, Eq. (71) leads to
\begin {equation}
\phi = \big[\mu (|\alpha| + |\beta|) + i(|\alpha| - |\beta|)\big]\ \theta_1
+ \big[|\alpha| + |\beta| + i\nu (|\alpha| - |\beta|)\big]\ \theta_2
\end {equation}
Identifying this expression with Eq. (48), we obtain
\begin {equation}
C_1 = \mu (|\alpha| + |\beta|) + i(|\alpha| - |\beta|)
\end {equation}
\begin {equation}
C_2=|\alpha| + |\beta| + i\nu (|\alpha| - |\beta|)
\end {equation}
Separating the real part from the imaginary part in these last two
equations, we obtain a system of four equations which can be solved 
with respect to $|\alpha|$, $|\beta|$, $\mu$ and $\nu$. It follows 
that for a given physical wave function $\phi$, these results fix 
univocally the reduced action. 

Note that in this reasoning, we have also eliminated the two 
superfluous degrees of freedom. The first one has been eliminated 
by discarding the global phase $\exp\left(i{{a+b}\over 2 }\right)$ 
meaning that we have chosen 
$\,\Omega = \exp\left(-i{{a+b}\over 2 }\right),\,$ 
and the second one in the judicious choice of the value of the 
integration constant $\lambda$.

\vskip0.5\baselineskip
\noindent
{\bf 5.4\ \ Microstates and bound states}
\vskip0.5\baselineskip

For bound states $(|\alpha|=|\beta|)$, Eq. (71) reduces to
\begin {equation}
\phi = 2\, |\alpha| A\ \cos{\left[\arctan{\left({ \theta_1 + 
\nu \theta_2} \over {\mu \theta_1 + \theta_2}\right)}   
\right]}
\end {equation}
Using expression (43) for $A$, this last equation becomes
\begin {equation}
\phi = 2\,|\alpha| (\mu \theta_1 + \theta_2 ) \ . 
\end {equation}
Identifying now this expression with (48), we obtain
\begin {equation}
C_1 = 2\, |\alpha| \mu \ ,\ \ \ \ \ \ \ \ \ \ \ \ C_2 = 2\, |\alpha|
\end {equation}
which leads to
\begin {equation}
\mu = {C_1 \over C_2} \ . 
\end {equation}
It is clear that the initial conditions (49) and (50) of the 
Schr\"odinger wave function do not allow to fix the value of $\nu$. 
According to Floyd's proposal \cite{F2} for which time 
parameterization is given by Jacobi's theorem : 
$t-t_0 = \partial S_0 / \partial E $,
Eqs. (41) and (77) mean that we obtain, for a given physical state 
$\phi$, a time dependent family of trajectories which can be 
specified by the different values of $\nu$. Hence, we can assert 
that $\nu$ plays the role of a hidden variable and specifies, 
for the same state $\phi$, the different microstates not detected 
by the Schr\"odinger wave function.

Note that we can also explain these results in the following manner. 
Substituting in (27) the function $A$ by its expression (39), we 
obtain
\begin {equation}
\phi = D \ 
\left({\partial S_{0} \over \partial x  }\right)^{-1/ 2 }\  
\cos\left( {S_{0} \over\hbar } + {{a-b}\over 2} \right) \ , 
\end {equation}
where we have discarded the unimportant phase factor 
$
\exp {\left( i{{a+b} \over 2}\right)}
$
and set $D=2\vert\alpha\vert k$. Integrating this differential 
equation gives
\begin {equation}
S_{0} = \hbar\ \arctan{\left({1 \over \hbar} \int {dx\over{(\phi / D)^2 }}
  + H \right)} + \hbar{{b-a}\over 2 } \ ,
\end {equation}
where we have written explicitly the integration constant $\,H\,$  
which arises by calculating the integral appearing in the right 
hand side of this last equation. The presence of this arbitrary 
constant $\,H\,$ explains the fact that for any physical state 
$\phi$, some freedom subsists in the choice of the reduced action 
$S_0$, giving rise to the existence of microstates for bound 
states.

In conclusion, the Schr\"odinger wave function is not an exhaustive 
description of non-relativistic systems. The quantum Hamilton-Jacobi 
equation is more fundamental. Hence, we confirm the finding of Floyd 
\cite{F5} who also showed that trajectory representation described 
microstates for bound states and not for unbound states. Finally, we 
would like to add that the problem of microstates was also 
investigated  by Carroll \cite{Carr} and by Faraggi-Matone-Bertoldi 
\cite{FM2,BFM}.

\vskip0.5\baselineskip
\noindent
{\bf 5.5\ \ The hidden antisymmetric two-tensor }
\vskip0.5\baselineskip

In three dimensions, the continuity equation (25) or (37) indicates
that we can write
\begin {equation}
 A^{2} \vec{\nabla  } S_0 = \vec{\nabla }\times \vec{B  } \ ,  
\end {equation}
where $\vec{B }$ is a vector field for which we can associate the 
two-tensor 
\begin {equation}
 F^{ij} = \partial^{i}B^{j} - \partial^{j}B^{i} 
\end {equation}
such that
\begin {equation}
A^{2} \partial_{i} S_0 = {1 \over 2}\epsilon_{ijk} F^{jk} \ ,  
\end {equation}
$i$, $j$ and $k$ represent the indices $(x,y,z)$ and $\epsilon_{ijk}$
is the usual Levi-Civita antisymmetric tensor. The continuity equation 
takes the form
\begin {equation}
 \epsilon_{ijk} \partial^{i}F^{jk} = 0  \ ,  
\end {equation}
with $F^{ij}$ being a hidden antisymmetric two-tensor field which may 
play, as suggested by Bertoldi-Faraggi-Matone \cite{BFM}, an important 
role in the understanding of the quantum origin of fundamental 
interactions.

\vskip0.5\baselineskip
\noindent
{\bf 5.6\ \ Velocity}
\vskip0.5\baselineskip
The last point which we will discuss in this section concerns the 
velocity. At first sight, according to the form (19) of the wave 
function, one may think that we can associate in any point of the 
trajectory two values for the velocity. In fact, the two components 
$\exp\left(+{i\over\hbar}S_0\right)$ and
$\exp\left(-{i\over\hbar}S_0\right)$ 
describe propagation of waves in opposite directions. In this way,
and interpreting the probability current as a sum of two currents 
(Eq. (31)), we can associate a motion of the particle to each of 
these components and define then at any point of the trajectory 
two velocities by using the hydrodynamic approach \cite{Bohm,LdB}
\begin {equation}
\vec{v}_{+} = {\vec{\jmath}_{+} \over {\cal A}^2 }\ , \ \ \ \ \ \ \ \ \ 
\vec{v}_{-}= {\vec{\jmath}_{-} \over {\cal A}^2 }  
\end {equation}
where $\,{\cal A}\,$ is the amplitude of the wave function
\begin {equation}
{\cal A} =A \ \left[ |\alpha|^2 + |\beta|^2 + 2\,|\alpha||\beta|
\cos{\left( {2\,S_0 \over \hbar} + a - b \right)}\right]^{1/2} 
\end {equation}
However, in the one-dimensional case with a constant potential, Floyd
showed \cite{F8} that it was possible for two waves propagating in opposite 
directions, to synthesize a single wave propagating only in one 
direction. This result although obtained for a particular case, 
indicates that the hydrodynamic approach is not appropriate to define 
correctly the particle velocity. This is different from the ideas
developed by Brown and Hiley \cite{BH} who consider that the particle 
velocity is the same to the current velocity  and assert
that the use of the classical canonical theory to define the
momentum is totally unnecessary. Without appealing to the hydrodynamic 
approach, Floyd showed \cite{F2} from the quantum Hamilton-Jacobi equation 
and Jacobi's theorem
$(\partial S_0 / \partial E = t-t_0)$, 
that in one dimension the velocity is given by relation (10).
In the same spirit, Carroll showed in Ref. \cite{Carr} that the current 
velocity could not be identified with the particle velocity.

\vskip0.5\baselineskip
\vskip0.5\baselineskip

\noindent
{\bf 6\ \ CONCLUSION}

\vskip0.5\baselineskip

In the three-dimensional space, it is shown in this paper that the
wave function, whether real or complex, has the unified form (19)
which leads to the same quantum Hamilton-Jacobi equation (26) or
(38), and the functions $A$ and $S_0$ are related by the same
continuity equation. The problem of the vanishing value of the 
conjugate momentum for real wave functions appearing in Bohm's 
theory is solved by the fact that the reality of the wave function 
is not expressed by $S_0=cst.$ but by 
$\vert\alpha\vert=\vert\beta\vert$.
Let us insist on the fact that the quantum Hamilton-Jacobi equation
obtained here is fundamentally different from the usual one because
the reduced action $S_0$ is related to the wave function by (19).

We have also proposed in one dimension a new form of the general 
solution of the quantum Hamilton-Jacobi equation and shown in the 
bound state case that there are  microstates not detected by the 
Schr\"odinger wave function. In three dimensions, 
we have seen that there is a hidden antisymmetric two-tensor field 
underlying quantum mechanics, recently introduced by 
Bertoldi-Faraggi-Matone. 

\vskip\baselineskip

\noindent
{\bf ACKNOWLEDGMENTS}
\vskip0.5\baselineskip
The author would like to thank Dr. K.~Adel for useful discussions.

\vskip\baselineskip

\end {document}